\documentclass[12pt, prd, showpacs]{revtex4}
\usepackage{amssymb}
\usepackage{amsmath}

\input{tcilatex}

\begin{document}

\title{General properties of the electric Penrose process }
\author{O. B. Zaslavskii}
\affiliation{Department of Physics and Technology, Kharkov V.N. Karazin National
University, 4 Svoboda Square, Kharkov 61022, Ukraine}
\email{zaslav@ukr.net }

\begin{abstract}
We consider the Penrose process with the charged particles in the
Reissner-Nordstr\"{o}m (RN) background. Let parent particle 0 decay to
particles 1 and 2. With the assumption that all three particles move in the
equatorial plane, the exact formulas for characteristics of particles 1 and
2 in terms of those of particle 0 are derived. We concentrate on scenarios
in which particle 1 and 2 are ejected along the trajectory of particle 0. It
is shown that such scenarios correspond to the extrema of energies $E_{1}$
or $E_{2}$ of daughter particles with respect to the angular momentum $L_{1}$
or $L_{2}$. We derive bounds on the values of angular momenta $L_{1}$ and $%
L_{2}$. We give classification of these scenarios and discuss their
properties including decay in the near-horizon region. We find that the
maximum of efficiency is achieved on the horizon for some of these scenarios
but not for all of them and with additional constraints on particle
parameters. The results are reformulated in terms of velocities of daughter
particles in the center of mass frame. The approach is applicable also to
collisional Penrose process, in which a combination of particles 1 and 2 is
considered as one effective particle. If the mass of particle 0 $%
m_{0}\rightarrow \infty $, the situation corresponds to the Ba\~{n}%
ados-Silk-West effect, the results agree with the ones known in literature
before. In addition, we consider special cases when decay occurs in the
turning point for one or all three particles. The formalism developed in
this work has a model-independent character and applies not only to the RN
metric.
\end{abstract}

\maketitle

\section{Introduction}

The Penrose process (hereafter PP) means that a parent particle 0 decays to
two daughter ones 1 and 2 in such a way that one of them (say, 1) has a
negative energy whereas particle 2 returns to infinity with an energy bigger
than an initial energy $E_{0}$ of particle 0. This process becomes possible
if in a space-time there exists the region (called ergoregion or ergosphere)
inside which $g_{\mu \nu }\xi ^{\mu }\xi ^{\nu }>0$, where $g_{\mu \nu }$
are metric coefficients and $\xi $ is the Killing vector responsible for
time translation (the signature of a metric is chosen to be (-,+,+,+)) \cite%
{pen}, \cite{pen2}. Meanwhile, a counterpart of the original PP in the
Reissner-Nordstr\"{o}m (RN) background was found due to the properties of
particle dynamics even in spite of the absence of such an ergoregion \cite%
{ruf}, \cite{den}. Instead, in this metric negative energies are possible in
some region (called generalized ergoregion) whose border depends on the
electric charge, angular momentum and mass of a particle.

More recently, we have seen a new wave of interest to the PP in new
contexts. This includes the collisional version of it \cite{schrev},
confined one \cite{conf}, \cite{myconf}, the relation between the so-called
Ba\~{n}ados-Silk-West (BSW) effect \cite{ban} and the PP, the PP for
spinning particles \cite{ext}, binaries \cite{bin}, \cite{pp2bh}, the PP in
the background of the Vaidya space-time \cite{ppv}. In case of getting
formally unbounded \ energies, the PP is called super-Penrose (SPP) effect
(see, e.g. \cite{is} and references therein). The subtleties connected with
the difference between power and efficiency of the PP was discussed in \cite%
{trade}.

As a rule, the investigations of the BSW effect were based on the careful
analysis of the vicinity of the horizon. If we characterize the proximity to
the horizon by the value of the lapse function $N$, this implies that there
is a small parameter $N\ll 1$. Meanwhile, there is another approach that is
based on exact formulas describing decay. For neutral particles this was
realized in \cite{centr}. The meaningful difference between (i) collisions
near a rotating black hole and (ii) a static charged one consists in that in
case (i) there is an upper bound forbidding the SPP near black holes but
there is no such a bound for (ii) \cite{rn}, \cite{fh}.

The emphasis in \cite{ruf}, \cite{den} and the main part of subsequent works
of this trend (see \cite{dad3} for review) was made on the properties of
negative energy states for a given particle in the presence of the
electromagnetic field for a concrete metric \cite{rufkerr}. Meanwhile, we
are going to focus on another aspect connected with the relation between an
initial state and its products after decay. We elucidate how characteristics
of particle 0 and those of particles 1 and 2 are related with each other,
which scenarios are possible in principle and what are their properties.

Although for a realistic process in astrophysics it is necessary to take
into account both rotation and electromagnetic field that includes not only
the electric but also the magnetic one \cite{dad1}, \cite{dad2}, \cite{luca}%
, \cite{fifty}, \cite{uniform} (see also a useful list of references on the
subject in \cite{win}), it looks reasonable at the first stage to consider a
simplified problem and separate two different factors - rotation and
electric charge. In the previous work \cite{eq} we gave full classification
of possible scenarios for the particle decay and PP with neutral particles
in the rotating \ background, provided all processes occur within the
equatorial plane. In the present work, we consider a similar problem for
electrically charged particles in the static black hole metric with the
charge, i.e. the Reissner-Nodstr\"{o}m (RN) one. It is worth noting that a
similar problem was considered recently in \cite{ju} but for weakly charged
black holes only and one scenario whereas in our approach we take into
account this charge exactly and give classification of all possible
scenarios. What is more important, our results are qualitatively different
(see discussion below).

Although the main area of application is the RN metric, practically all
formulas do not use its particular form. This means that our results apply
to the wide class of spherically symmetric static metrics described by eq. (%
\ref{met}) below. Although they do not apply to axially symmetric stationary
ones directly, we hope that general idea is useful for them as well, even
under the presence of the electromagnetic field. The reason is that our
scheme is quite general. (i) We do not restrict ourselves by a particular
scenario in which decay occurs in a turning point of radial motion but build
classification scheme of all possible scenarios, (ii) we find explicitly the
expression for a velocity of fragments after decay, (iii) we trace the
relation between the electromagnetic version \cite{dad0} of the Wald
approach \cite{wald} and particle dynamics of decay, thus relating the
velocities and masses, (iv) we find explicitly the angular momenta of
daughter particles. We will see that the roles of rotation and electric
charge are very different and some scenarios of the PP forbidden inside the
ergosphere of the rotating metric \cite{eq} are allowed in the generalized
ergosphere of the static charged metric. Thus the present work and the
previous one \cite{eq} constitute the mutually complementary pair that we
hope to combine in a unified picture later.

A separate question is the dependence of the efficiency of the process on
the position of the decay point. There is a popular aproach in which this
pint is chosen as close to the horizon as possible, with point of decay
coinciding with a turning point of particle 0 (see, e.g. Sec. 5.1 of \cite%
{dad1}). Meanwhile, we show that this is not always so and depends strongly
on a type of scenario the list of which we discuss carefully case by case.
Bearing in mind future application to astrophysical process, this seems to
be important enough since it alows to describe different cases without
additional arificial assumptions about a particular type of decay.

The paper is organized as follows. In Sec. \ref{setup} we list equations of
particle motion in the RN background. In Sec. \ref{dec} we give the exact
formulas that related characteristics of particle 0 and particles 1, 2. In
Sec. \ref{class} we suggest classification of scenarios when particle 0 is
ingoing. In Sec. \ref{along} we make emphasis on a type of decay when
particles 1 and 2 are ejected along the trajectory of particle 0. In Sec. %
\ref{massless} we consider a special case than an escaping particle is
massless. In Sec. \ref{proper} we consider some properties of scenario I
including bounds on the energy and angular momentum of an escaping particle
and decay in the near-horizon region. In\ Sec. \ref{decayturn} we discuss
the situation when the point of decay coincides with the turning point of
radial motion for particle 0, particle 2 or with a common turning point of
all three particles. In Sec. \ref{thres} we derive the conditions necessary
for the PP process in the scenarios under discussion. In Sec. \ref{vel} we
reformulate our result kinematically, using the velocities of particles and
their Lorentz gamma-factors. In Sec. \ref{inv} we consider the scenario in
which particle 0 is outgoing. In Sec. \ref{prox} we consider the conditions
under which the maximum of the efficiency is achieved on the horizon
depending on scenario. In Sec. \ref{comp} we summarize the main features of
the considered scenarios with short comments. In Sec. \ref{discr} we compare
our results with some other ones recently published and explain the origin
of discrepancy. In Sec. \ref{bsw} we display the relation between our
approach and that used in literature for the description of the BSW effect
before. In Sec. \ref{concl} we give the summary of main results. In Appendix
we derive some useful inequalities relevant for the PP near the horizon.

We use system of units in which fundamental constants $G=c=1$.

\section{Charged particles in the metric of Reissner-Nordstr\"{o}m black
hole \label{setup}}

We consider the metric that has the form%
\begin{equation}
ds^{2}=-dt^{2}f+\frac{dr^{2}}{f}+r^{2}d\omega ^{2},  \label{met}
\end{equation}%
with $d\omega ^{2}=d\theta ^{2}+\sin ^{2}\theta d\phi ^{2}.$ We will mainly
deal with the RN metric, then $f\equiv N^{2}=1-\frac{2M}{r}+\frac{Q^{2}}{%
r^{2}}=\frac{(r-r_{+})(r-r_{-})}{r^{2}}$. Here, $M$ is a black hole mass, $Q$
is its electric charge (for definiteness $Q>0$), $r_{+}=M+\sqrt{M^{2}-Q^{2}}$
being the event horizon radius, $r_{-}$ $=M-\sqrt{M^{2}-Q^{2}}$ the inner
horizon radius. Meanwhile, basic results are extendable to an arbitrary $f$
outside the event horizon.

Let us consider motion of a charged particle in this space-time without
additional forces. Then, there are two integrals of motion. As the metric
does not depend on time and angle $\phi $, the energy $E$ and angular
momentum $L$ are conserved. Motion occurs within a plane which we choose to
be $\theta =\frac{\pi }{2}$. Then, equations of motion read (dot denotes
derivative with respect to the proper time $\tau $)%
\begin{equation}
\dot{t}=\frac{X}{mf}\text{,}  \label{td}
\end{equation}%
\begin{equation}
X=E-q\varphi \text{,}  \label{X}
\end{equation}%
\begin{equation}
\dot{\phi}=\frac{L}{mr^{2}}\text{,}
\end{equation}%
\begin{equation}
m\dot{r}=\sigma P\text{,}
\end{equation}%
\begin{equation}
P=\sqrt{X^{2}-N^{2}\tilde{m}^{2}}\text{,}  \label{P}
\end{equation}%
\begin{equation}
\tilde{m}^{2}=m^{2}+\frac{L^{2}}{r^{2}}.  \label{mdef}
\end{equation}%
Here, $q$ is a particle's charge, $\varphi =\frac{Q}{r}$ is the Coulomb
potential of the RN metric , $\sigma =\pm 1$. The forward-in-time condition $%
\dot{t}>0$ requires%
\begin{equation}
X\geq 0\text{,}  \label{ft}
\end{equation}%
outside the horizon $X>0$.

From $P^{2}\geq 0$ we have a more tight condition than (\ref{ft}): 
\begin{equation}
X\geq \tilde{m}N.  \label{xmn}
\end{equation}

\section{Decay to two particles: general scenario\label{dec}}

Let in some point $r=r_{d}$ particle 0 decay to particles 1 and 2. For
simplicity, we assume that all three particles move within the same plane.
In the point of decay the conservation laws give us%
\begin{equation}
E_{0}=E_{1}+E_{2},  \label{e}
\end{equation}

\begin{equation}
L_{0}=L_{1}+L_{2}\text{,}  \label{L}
\end{equation}%
\begin{equation}
q_{0}=q_{1}+q_{2}\text{.}  \label{q}
\end{equation}%
It follows from (\ref{e}) - (\ref{q}) that%
\begin{equation}
X_{0}=X_{1}+X_{2}.  \label{x12}
\end{equation}%
The conservation of the radial component of momentum reads%
\begin{equation}
\sigma _{0}P_{0}=\sigma _{1}P_{1}+\sigma _{2}P_{2}\text{.}  \label{P0}
\end{equation}%
Let particle 0 move with decreasing $r$, so $\sigma _{0}=-1$. It is clear
that combination $\sigma _{1}=\sigma _{2}=+1$ is impossible since this would
contradict eq. (\ref{P0}). For definiteness, we assume that particle 1 moves
after decay with $\sigma _{1}=-1$. Moreover, we imply that in the frame
comoving with particle 0 (which is the center of mass frame of particles 1
and 2 in the point of decay), particle 1 moves in the inward direction as
well. Meanwhile, $\sigma _{2}$ can have any sign. In the center of mass
frame particle 2 moves in the outward direction with some velocity $v_{2}$.
However, if we pass to the frame comoving with particle 0 that moves in the
inward direction with the velocity $V_{2}$ in the static frame, both cases
are possible: $\sigma _{2}=-1$ or $\sigma _{2}=+1$ depending on which value
of the velocity is bigger (below we will discuss this issue in more detail).

Then we have%
\begin{equation}
P_{0}=P_{1}-\sigma _{2}P_{2}\text{.}  \label{ppp}
\end{equation}%
Solving equations of motion, we obtain%
\begin{equation}
X_{1}=\frac{X_{0}}{2\tilde{m}_{0}^{2}}\tilde{b}_{1}-\delta \frac{P_{0}\sqrt{%
\tilde{d}}}{2\tilde{m}_{0}^{2}}\text{,}  \label{e1}
\end{equation}%
\begin{equation}
X_{2}=\frac{1}{2\tilde{m}_{0}^{2}}\left( X_{0}\tilde{b}_{2}+P_{0}\delta 
\sqrt{\tilde{d}}\right) ,  \label{e2}
\end{equation}

\begin{equation}
P_{1}=\left\vert \frac{P_{0}\tilde{b}_{1}-\delta X_{0}\sqrt{\tilde{d}}}{2%
\tilde{m}_{0}^{2}}\right\vert \text{.}  \label{p1}
\end{equation}%
\begin{equation}
P_{2}=\left\vert \frac{P_{0}\tilde{b}_{2}+\delta X_{0}\sqrt{\tilde{d}}}{2%
\tilde{m}_{0}^{2}}\right\vert .  \label{p2}
\end{equation}

Here, $\delta =\pm 1$,%
\begin{equation}
\tilde{b}_{1,2}=\tilde{m}_{0}^{2}+\tilde{m}_{1,2}^{2}-\tilde{m}_{2,1}^{2}%
\text{,}
\end{equation}%
\begin{equation}
\tilde{d}=\tilde{b}_{1}^{2}-4\tilde{m}_{0}^{2}\tilde{m}_{1}^{2}=\tilde{b}%
_{2}^{2}-4\tilde{m}_{0}^{2}\tilde{m}_{2}^{2}\text{.}
\end{equation}%
For what follows, it is also convenient to introduce also corresponding
quantities without tilde:%
\begin{equation}
b_{1,2}=m_{0}^{2}+(m_{1,2}^{2}-m_{2,1}^{2}),  \label{b}
\end{equation}%
\begin{equation}
d=b_{1}^{2}-4m_{0}^{2}m_{1}^{2}=b_{2}^{2}-4m_{0}^{2}m_{2}^{2}.  \label{d}
\end{equation}

Then, it can be obtained from (\ref{L}) that%
\begin{equation}
\tilde{b}_{1,2}=b_{1,2}+\frac{2L_{0}L_{1,2}}{g_{\phi }},  \label{btil}
\end{equation}%
\begin{equation}
\tilde{d}=d+4b_{1}\frac{L_{0}L_{1}}{r^{2}}-\frac{4L_{1}^{2}}{r^{2}}%
m_{0}^{2}-4\frac{L_{0}^{2}}{r^{2}}m_{1}^{2},  \label{dtil}
\end{equation}%
\begin{equation}
\tilde{d}=d+4b_{2}\frac{L_{0}L_{2}}{r^{2}}-\frac{4L_{2}^{2}}{r^{2}}%
m_{0}^{2}-4\frac{L_{0}^{2}}{r^{2}}m_{2}^{2}.  \label{dtil2}
\end{equation}

Eqs. (\ref{e1}) - (\ref{p2}) are the counterpart of equations (19), (20),
(26), (27) in \cite{centr} obtained there for rotating metrics. The
essential difference consists in the expression for $X$. Now it is given by (%
\ref{X}) and does not depend on $L$. Meanwhile, in \cite{centr} $X=E-\omega
L $, the $\omega $ is the metric coefficient responsible for rotation. It is
the fact that $L$ does not enter the expression for $X$ that allows some
scenarios which were forbidden in the rotating case (see below).

\subsection{Unconditional bounds on mass}

From the condition $\tilde{d}\geq 0$ one can deduce that%
\begin{equation}
\tilde{m}_{0}\geq \tilde{m}_{1}+\tilde{m}_{2}.  \label{m}
\end{equation}

The same condition requires that, according to (\ref{dtil}), (\ref{dtil2})%
\begin{equation}
L_{1,2}^{(-)}\leq L_{1,2}\leq L_{1,2}^{(+)}\text{,}  \label{Lminmsx}
\end{equation}%
where%
\begin{equation}
L_{1,2}^{(\pm )}=\frac{b_{1,2}L_{0}}{2m_{0}^{2}}\pm \frac{\sqrt{d}r_{d}}{%
2m_{0}^{2}}\tilde{m}_{0}\text{.}  \label{L+-}
\end{equation}

If one takes into account the definition (\ref{mdef}) and the conservation
law (\ref{L}), one can obtain by straightforward algebraic manipulations
that 
\begin{equation}
m_{0}\geq m_{1}+m_{2}  \label{mp}
\end{equation}%
as well.

One can also use the conservation law for four-momenta in the covariant form 
\begin{equation}
p_{0}^{\mu }=p_{1}^{\mu }+p_{2}^{\mu }\text{.}
\end{equation}

Then, taking the square and taking into account that for future-directed
four-vectors $p_{1\mu }p_{2}^{\mu }<0$, it is easy to obtain%
\begin{equation}
m_{0}^{2}\geq m_{1}^{2}+m_{2}^{2}\text{.}  \label{m22}
\end{equation}

This is just eq. (4.11) in \cite{dad1} and eq. (3.11) in \cite{dad2},
written in our notations. Meanwhile, eq. (\ref{mp}) is more tight than (\ref%
{m22}). It can be derived if one uses the property $p_{1\mu }p_{2}^{\mu
}<-m_{1}m_{2}$ which is more tight than $p_{1\mu }p_{2}^{\mu }<0$ used in 
\cite{dad1}, \cite{dad2}.

\section{Classification of scenarios with respect to parameters in general 
\label{class}}

Each scenario of decay can be characterized by the set $(\sigma _{2}$, $%
h_{2} $, $h_{1}$, $\delta )$, where%
\begin{equation}
h_{1,2}=sign(\tilde{b}_{1,2}N-2\tilde{m}_{1,2}X_{0}).  \label{h}
\end{equation}%
The quantities $h_{1}$ and $h_{2}$ arise in a natural way, when we sort out
combinations with different signs inside the absolute values in (\ref{p1})
and (\ref{p2}) and determine which term dominates. As a result, one obtains
the following types of scenarios:

I $(\sigma _{2}=+1$, $h_{2}=0$ or $h=+1$, $\delta =-1)$%
\begin{equation}
P_{1}-P_{2}=P_{0}\text{,}  \label{1}
\end{equation}%
\begin{equation}
P_{2}=\frac{(X_{0}\sqrt{\tilde{d}}-P_{0}\tilde{b}_{2})}{2\tilde{m}_{0}^{2}}%
\text{,}
\end{equation}%
\begin{equation}
P_{1}=\frac{P_{0}\tilde{b}_{1}+X_{0}\sqrt{\tilde{d}}}{2\tilde{m}_{0}^{2}},
\end{equation}%
\begin{equation}
X_{1}=\frac{X_{0}}{2\tilde{m}_{0}^{2}}\tilde{b}_{1}+\frac{P_{0}\sqrt{\tilde{d%
}}}{2\tilde{m}_{0}^{2}}\text{,}
\end{equation}%
\begin{equation}
X_{2}=\frac{1}{2\tilde{m}_{0}^{2}}\left( X_{0}\tilde{b}_{2}-P_{0}\sqrt{%
\tilde{d}}\right) .
\end{equation}%
II $(\sigma _{2}=-1$, $h_{2}=-1$, $\delta =-1)$%
\begin{equation}
P_{1}+P_{2}=P_{0},  \label{2}
\end{equation}

\begin{equation}
P_{2}=\frac{(P_{0}\tilde{b}_{2}-X_{0}\sqrt{\tilde{d}})}{2\tilde{m}_{0}^{2}}%
\text{,}
\end{equation}%
\begin{equation}
P_{1}=\frac{P_{0}\tilde{b}_{1}+X_{0}\sqrt{\tilde{d}}}{2\tilde{m}_{0}^{2}}%
\text{,}
\end{equation}%
\begin{equation}
X_{1}=\frac{X_{0}}{2\tilde{m}_{0}^{2}}\tilde{b}_{1}+\frac{P_{0}\sqrt{\tilde{d%
}}}{2\tilde{m}_{0}^{2}}\text{,}
\end{equation}%
\begin{equation}
X_{2}=\frac{1}{2\tilde{m}_{0}^{2}}\left( X_{0}\tilde{b}_{2}-P_{0}\sqrt{%
\tilde{d}}\right) \text{.}
\end{equation}

III $(\sigma _{2}=-1$, $h_{1}=0$ or $h=-1$, $\delta =+1)$

\begin{equation}
P_{2}=\frac{X_{0}\sqrt{\tilde{d}}+P_{0}\tilde{b}_{2}}{2\tilde{m}_{0}^{2}}%
\text{,}
\end{equation}%
\begin{equation}
P_{1}=\frac{P_{0}\tilde{b}_{1}-X_{0}\sqrt{\tilde{d}}}{2\tilde{m}_{0}^{2}}%
\text{,}
\end{equation}%
\begin{equation}
X_{1}=\frac{X_{0}}{2\tilde{m}_{0}^{2}}\tilde{b}_{1}-\frac{P_{0}\sqrt{\tilde{d%
}}}{2\tilde{m}_{0}^{2}}\text{,}
\end{equation}%
\begin{equation}
X_{2}=\frac{1}{2\tilde{m}_{0}^{2}}\left( X_{0}\tilde{b}_{2}+P_{0}\sqrt{%
\tilde{d}}\right) .
\end{equation}

Scenarios II and III can be obtained from each other by interchange of
particles 1 and 2. In this sense, they are equivalent, so for definiteness
we will consider scenario II.

\section{Types of scenarios and their meaning}

Formulas of the previous Section describe general relations between a parent
particle and daughter ones. Meanwhile, for physical purposes, we are
interested in more concrete scenarios. The most simple and popular one
consists in such a decay of particle 0 to 1 and 2 that all three particles
are located in the corresponding turning points. Meanwhile, in any real
process, one cannot guarantee the validity of this condition and, therefore,
one is led to considering more expanded set of possible scenarios. In the
first place, this concerns the situation when all three particles are moving
and the products of decay are ejected along the trajectory. This is of
special interest since this leads to the most effective (for a given \ state
if particle 0) kind of process. (This will be proven in the next Section.)
In turn, here there are different cases depending on how a daughter particle
move after decay, whether it moves towards a black hole or towards infinity.
There are also intermediate cases when some of particles (but not all of
them) are in their turning points.

The key ingredient of the Penrose process is the existence of the ergoregion
(in the standard PP) or generalized ergoregion (in the electric version of
PP), where particle energy can be, in principle, negative. In this respect
it was noticed earlier \cite{eq} that for a rotating black hole some kinds
of decay, which seem to be compatible with the existence of negative states
for an individual particle, were forbidden in the ergoregion. It is of
interest to elucidate, whether or not this happens for charged particles in
the generalized ergoregion of a static charged black hole.

More precisely, we consider the following scenarios.

I: particle 2 moves in the outward direction, particle 1 moves in the inward
one. Further, particle 1 can fall in a black hole while particle 2 can
escape.

II: particles 1 and 2 move in the same direction as particle 0, i.e. inward
direction.

In scenarios I and II it is implied that before decay particle 0 moved in
the inward direction, so it follows from the momentum conservation that both
particles 1 and 2 cannot move after decay in the outward direction.

B ("bounce"). Particle 0 bounces back from the potential barrier, moves in
the outward direction and decays afterwards. Preliminarily, this scenario
gives us the most efficient possibility when a daughter particle moves in
the same direction as particle 0. Formally, this is similar to scenario II
but in scenario II both particles move towards a black hole whereas in
scenario B they start to move to infinity.

For completeness, we take into account also the existence of a turning point.

TP2: point of decay of particle 0 coincides with the turning point of decay
of particle 2.

TP0: point of decay of particle 0 coincides with its turning point but not
with the turning points of particles 1 and 2.

TP3: decay occurs in the common point of decay for all three particles.

Below, we consider all these scenarios case by case. Bearing in mind
potential applications to processes near a black hole, of primary interest
is the near-horizon behavior of particles in these processes. The results
are collected in Sec. \ref{comp} below where we also indicate near-horizon
properties. A special questin about the role of a horizon in achieving the
maximum of efficiency is considered in Sec. \ref{prox}.

\section{Ejection along trajectory: maximization/minimization of energy and
classification of scenarios \label{along}}

Now, we consider a special type of decay in which products of decay are
ejected along the trajectory of the original particle.

In doing so,

\begin{equation}
\frac{P_{1}}{P_{0}}=\frac{L_{1}}{L_{0}}\text{,}  \label{ej}
\end{equation}%
whence 
\begin{equation}
(P_{0}\tilde{b}_{1}-\delta X_{0}\sqrt{\tilde{d}})L_{0}-2\tilde{m}%
_{0}^{2}L_{1}P_{0}=0\text{.}
\end{equation}

For definiteness, we assumed that signs of angular momenta of particles 0
and 1 coincide.

The interest to such a type of trajectory is motivated by the fact that it
is the most efficient process (with other parameters of scenario fixed). The
particle that moves in the same direction as particle 0 acquires the maximum
possible energy, the other one that moves in the opposite direction has the
minimum one. To show this, one can consider, say, energy $E_{1}$ as a
function of $L_{1}$:%
\begin{equation}
E_{1}=\frac{X_{0}}{2\tilde{m}_{0}^{2}}\tilde{b}_{1}-\delta \frac{P_{0}\sqrt{%
\tilde{d}}}{2\tilde{m}_{0}^{2}}+q_{1}\varphi \text{.}  \label{e1d}
\end{equation}%
In the case of extremum, we have%
\begin{equation}
\frac{\partial E_{1}}{\partial L_{1}}=0\text{.}  \label{dE}
\end{equation}%
After taking the square and performing a number of algebraic manipulations,
we obtain just expressions for $X_{1}$ and $X_{2}$ that can be obtained from
(\ref{ej}). We list them below explicitly.

There are two different possible scenarios here. We consider them separately.

\subsection{Scenario I, $\protect\sigma _{2}=+1$, $\protect\delta =-1$, $%
h_{2}=+1$}

Remarkably, it turns out that after somewhat long algebraic manipulation,
one can get rid off tilted quantities and express the results in a rather
simple form in terms of quantities without tilde:%
\begin{equation}
X_{2}=X_{0}\frac{b_{2}}{2m_{0}^{2}}-\frac{\sqrt{d}\sqrt{%
X_{0}^{2}-m_{0}^{2}N^{2}}}{2m_{0}^{2}},  \label{e2I}
\end{equation}%
\begin{equation}
X_{1}=X_{0}\frac{b_{1}}{2m_{0}^{2}}+\frac{\sqrt{d}\sqrt{%
X_{0}^{2}-m_{0}^{2}N^{2}}}{2m_{0}^{2}}.  \label{e1I}
\end{equation}%
\begin{equation}
P_{2}=\frac{P_{0}}{2m_{0}^{2}\sqrt{X_{0}^{2}-m_{0}^{2}N^{2}}}(X_{0}\sqrt{d}%
-b_{2}\sqrt{X_{0}^{2}\mathcal{-}m_{0}^{2}N^{2}})\text{,}  \label{P2f}
\end{equation}%
\begin{equation}
P_{1}=\frac{P_{0}}{2m_{0}^{2}\sqrt{X_{0}^{2}-m_{0}^{2}N^{2}}}(b_{1}\sqrt{%
X_{0}^{2}-m_{0}^{2}N^{2}}+X_{0}\sqrt{d}).  \label{P1f}
\end{equation}

In general in eqs. (\ref{e1}) - (\ref{p2}) one of angular momenta (say, $%
L_{1}$) was a free parameter. However, now this is not so since there is an
additional constraint (\ref{ej}) that leads to

\begin{equation}
L_{2}=\frac{L_{0}}{2m_{0}^{2}}(b_{2}-\frac{\sqrt{d}}{\sqrt{%
X_{0}^{2}-N^{2}m_{0}^{2}}}X_{0})\text{,}  \label{L2I}
\end{equation}%
\begin{equation}
L_{1}=\frac{L_{0}}{2m_{0}^{2}}(b_{1}+\frac{\sqrt{d}X_{0}}{\sqrt{%
X_{0}^{2}-m_{0}^{2}N^{2}}})\text{.}  \label{L1I}
\end{equation}%
One can check that eqs. \ (\ref{P}) and (\ref{ej}) do hold for each particle
1 and 2.

The requirement $d\geq 0$ entails the inequality%
\begin{equation}
m_{0}\geq m_{1}+m_{2},  \label{m0}
\end{equation}%
similar to (\ref{m}) and coinciding with (\ref{mp}).

It follows from $P_{2}\geq 0$ that

\begin{equation}
b_{2}\sqrt{X_{0}^{2}-m_{0}^{2}N^{2}}\leq X_{0}\sqrt{d}\,,  \label{rad1}
\end{equation}%
whence%
\begin{equation}
X_{0}\leq \frac{Nb_{2}}{2m_{2}}.  \label{cr1}
\end{equation}%
As the expression inside the square root should be non-negative, $X_{0}\geq
m_{0}N$. It holds automatically, if (\ref{xmn}) does so. Combining the above
inequalities and (\ref{xmn}), we have 
\begin{equation}
\tilde{m}_{0}N\leq X_{0}\leq \frac{Nb_{2}}{2m_{2}}.  \label{ineq1}
\end{equation}%
For $L_{0}=0$ (so $\tilde{m}_{0}=m_{0}$) this is consistent with (\ref{cr1})
automatically since%
\begin{equation}
b_{2}\geq 2m_{0}m_{2}  \label{b2}
\end{equation}%
due to (\ref{m0}).

\subsection{Scenario II}

$\sigma _{2}=-1\,,$ $\delta =-1$%
\begin{equation}
P_{1}+P_{2}=P_{0\text{,}}  \label{p+}
\end{equation}

\begin{equation}
L_{1}=\frac{L_{0}}{2m_{0}^{2}}(b_{1}+\frac{\sqrt{d}X_{0}}{\sqrt{%
X_{0}^{2}-m_{0}^{2}N^{2}}})\text{,}  \label{L2II}
\end{equation}

\begin{equation}
L_{2}=\frac{L_{0}}{2m_{0}^{2}}(b_{2}-\frac{\sqrt{d}X_{0}}{\sqrt{%
X_{0}^{2}-m_{0}^{2}N^{2}}}),  \label{L1II}
\end{equation}

\begin{equation}
X_{1}=\frac{X_{0}b_{1}}{2m_{0}^{2}}+\frac{\sqrt{d}}{2m_{0}^{2}}\sqrt{%
X_{0}^{2}-m_{0}^{2}N^{2}},  \label{IIe1}
\end{equation}%
\begin{equation}
X_{2}=\frac{b_{2}X_{0}}{2m_{0}^{2}}-\frac{\sqrt{d}}{2m_{0}^{2}}\sqrt{%
X_{0}^{2}-m_{0}^{2}N^{2}}.  \label{IIe2}
\end{equation}%
\begin{equation}
P_{1}=\frac{P_{0}}{2m_{0}^{2}\sqrt{X_{0}^{2}-N^{2}m_{0}^{2}}}(b_{1}\sqrt{%
X_{0}^{2}-m_{0}^{2}N^{2}}+X_{0}\sqrt{d})\text{,}  \label{P2II}
\end{equation}%
\begin{equation}
P_{2}=\frac{P_{0}}{2m_{0}^{2}\sqrt{X_{0}^{2}-N^{2}m_{0}^{2}}}(b_{2}\sqrt{%
X_{0}^{2}-m_{0}^{2}N^{2}}-X_{0}\sqrt{d}).  \label{P1II}
\end{equation}%
It follows from $P_{2}\geq 0$ that%
\begin{equation}
X_{0}\geq \frac{b_{2}N}{2m_{2}}  \label{cr2}
\end{equation}%
and%
\begin{equation}
Nm_{2}\leq X_{2}\leq \frac{b_{2}}{2m_{0}^{2}}X_{0}\text{,}  \label{x22}
\end{equation}%
where we took into account (\ref{cr2}) and (\ref{b}), (\ref{d}).

It is worth noting that in scenario I $L_{2}$ has the sign opposite to that
of $L_{0}$, in scenario II they coincide.

If $Q\rightarrow 0$, we arrive at the Schwarzschild metric. Then, our
formulas agree with \cite{ob}. In this case, there is no the PP and the
energy of an escaping particle $E_{2}<E_{0}$ but $E_{2}\neq 0$.

\section{Massless case\label{massless}}

A special case arises when particle 2 is massless since condition (\ref{cr2}%
) is no longer valid if $m_{2}=0$. In turn, this means that scenario II
fails. This can be explained as follows. In general, scenario II is realized
when particle 2 is ejected in the outward direction in the frame comoving
with particle 0 (that is also the center of mass frame for particles 1 and
2). However, when we pass to the stationary frame, it is drifted in the
inward direction due to motion of particle 0 (see Sec. \ref{vel} below for
discussion of local Lorentz transformations). However, if particle 2 is
massless, it moves with a speed of light and inward motion of particle 0
cannot overcome outward motion of particle 2. As a result, particle 2 moves
outwardly that corresponds to scenario I, not II.

Therefore, we must use eqs. (\ref{e2I}), (\ref{P2f}), in which we put $%
m_{2}=0$ that gives us $b_{2}=d=m_{0}^{2}-m_{1}^{2}$. Thus we have%
\begin{equation}
X_{2}=\frac{b_{2}}{2m_{0}^{2}}(X_{0}-\sqrt{X_{0}^{2}-m_{0}^{2}N^{2}}),
\end{equation}%
\begin{equation}
P_{2}=\frac{P_{0}b_{2}}{2m_{0}^{2}\sqrt{X_{0}^{2}-m_{0}^{2}N^{2}}}(X_{0}-%
\sqrt{X_{0}^{2}\mathcal{-}N^{2}m_{0}^{2}})\text{.}
\end{equation}

In the horizon limit $N\rightarrow 0$, 
\begin{equation}
X_{2}\approx \frac{b_{2}N^{2}}{4X_{0}}\text{, }E\approx \frac{q_{2}Q}{r_{+}}+%
\frac{b_{2}N^{2}}{4X_{0}}\text{,}  \label{2m2}
\end{equation}%
\begin{equation}
X_{1}\approx X_{0}-\frac{b_{2}N^{2}}{4X_{0}}\text{, }E_{1}\approx E_{0}-%
\frac{q_{2}Q}{r_{+}}-\frac{b_{2}N^{2}}{4X_{0}}\text{.}
\end{equation}%
From another hand, if we put $m_{1}=0$, this does not give rise to any
difficulties since particle 1 moves inwardly in both frames, so scenario II
is valid as well.

\section{Properties of scenario I\label{proper}}

We are interested mainly in scenario I since it is this scenario in which
particle 2 can escape to infinity. First of all, we would like to stress the
difference between scenario I and its counterpart for processes with neutral
particles in the background of rotating black holes. In the latter case,
decay of type I is impossible in the ergoregion at all \cite{eq}. Meanwhile,
now it is allowed inside a generalized ergoregion. Below, we discuss some
details concerning just scenario I.

\subsection{Upper bound on the angular momentum}

For scenario I, there are restrictions (\ref{ineq1}). Both inequalities here
are consistent with each other, provided%
\begin{equation}
\tilde{m}_{0}\leq \frac{b_{2}}{2m_{2}}\text{.}  \label{mb}
\end{equation}%
Taking the square, we obtain the restriction on a value of $L_{0}$:%
\begin{equation}
L_{0}^{2}\leq L_{m}^{2}=\frac{dr_{d}^{2}}{4m_{2}^{2}}\text{.}  \label{L0}
\end{equation}%
Let $d\rightarrow 0$. This entails 
\begin{equation}
m_{0}=m_{1}+m_{2}.  \label{m012}
\end{equation}%
Then, scenario I is possible if $L_{0}\rightarrow 0$ only. According to (\ref%
{L2I}), (\ref{L1I}) this entails that $L_{1,2}\rightarrow 0$ as well, so all
three particles move radially. This scenario is realized in the confined
Penrose process \ \cite{conf}, \cite{myconf}. There is no similar
restriction on $L_{0}$ in scenario II.

\subsection{Upper bound on the energy of escaping particle}

It follows from (\ref{cr1}) that scenario I can be realized for $\left(
X_{0}\right) _{d}\leq \frac{N_{d}b_{2}}{2m_{2}}$ only. Then, from (\ref{e2I}%
) and (\ref{b}), (\ref{d}) we have 
\begin{equation}
Nm_{2}\leq X_{2}\leq \frac{b_{2}^{2}N}{4m_{0}^{2}m_{2}}.
\end{equation}

Thus for the escaping particle we have the upper bounds%
\begin{equation}
q_{2}\frac{Q}{r_{d}}+Nm_{2}\leq E_{2}\leq q_{2}\frac{Q}{r_{d}}+\frac{%
b_{2}^{2}N}{4m_{0}^{2}m_{2}}\text{.}  \label{up}
\end{equation}

\subsection{Near-horizon limit}

Assuming that $m_{2}\neq 0$, let us consider 
\begin{equation}
\left( X_{0}\right) _{d}=a\frac{N_{d}b_{2}}{2m_{2}}
\end{equation}%
where according to (\ref{ineq1}) the coefficient $a$ (pure number) obeys
inequalities 
\begin{equation}
a_{m}\leq a\leq 1\text{, }a_{m}=2\frac{m_{2}\tilde{m}_{0}}{b_{2}}
\end{equation}%
in combinations with inequalities (\ref{mb}), (\ref{L0}).

If $N_{d}\rightarrow 0$ while keeping $a$ constant, we obtain an approximate
equality:%
\begin{equation}
E_{2}\approx q_{2}\frac{Q}{r_{+}}+cm_{2}N_{d}\text{,}  \label{2hor}
\end{equation}%
where%
\begin{equation}
c=\frac{b_{2}^{2}a-\sqrt{d}\sqrt{a^{2}b_{2}^{2}-4m_{0}^{2}m_{2}^{2}}}{%
4m_{2}^{2}m_{0}^{2}}\text{.}
\end{equation}%
One can check that $\frac{\partial c}{\partial a}\leq 0$. Therefore, 
\begin{equation}
1\leq c\leq c(a_{m})=c_{m}\text{,}
\end{equation}%
where 
\begin{equation}
c_{m}=\frac{\tilde{m}_{0}(r_{+})b_{2}}{2m_{2}m_{0}^{2}}-\frac{\left\vert
L_{0}\right\vert \sqrt{d}}{2m_{2}m_{0}^{2}r_{+}}.  \label{c}
\end{equation}%
In particular, if $L_{0}=0$,%
\begin{equation}
c_{m}=\frac{b_{2}}{2m_{2}m_{0}}\geq 1
\end{equation}%
that agrees with (\ref{up}). If $L_{0}=L_{m}$, we obtain from (\ref{L0})%
\begin{equation}
c_{m}=1.
\end{equation}

If $a=1$, it follows that $c=1$ and we return to (\ref{up}) with equality
instead of inequality, $\ $so $E_{2}\approx q_{2}\frac{Q}{r_{+}}+m_{2}N_{d}.$

Thus the correction to the main term in $X_{2}$ has the order $N$. We arrive
at an important conclusion. If decay occurs near the horizon, the particle
that moves to infinity should be near-critical with $X_{0}=O(N_{d})$. Usual
particles cannot escape from the horizon at all.

In the massless case $m_{2}=0$ the correction has the order $N^{2}$
according to (\ref{2m2}).

For particle 1 we have in the near-horizon limit%
\begin{equation}
E_{1}\approx q_{1}\frac{Q}{r_{+}}+N_{d}(\frac{ab_{2}}{2m_{2}}-c)\text{.}
\label{1h}
\end{equation}

\section{Special type of scenario: decay in turning point\label{decayturn}}

There are special cases to be considered separately. They arise when decay
occurs in the turning point. (Hereafter, by turning point we imply for
brevity a turning point for radial motion, a particle can have in general an
angular moment and nonzero angular velocity.) It follows from the
conservation law (\ref{P0}) that the point of decay cannot coincide with the
turning point for two particles precisely. Either it is the turning point
for (i) only one particle or (ii) for all three at once. It follows from (%
\ref{P0}) or (\ref{P1f}), (\ref{P2II}) that in scenario I, case (i) is
possible if there exists a turning point for particle 2, not for particle 1.
Now, we will discuss (i) and (ii) case by case.

\subsection{Case (i), turning point for particle 2 (TP2)}

Let particle 2 arise in its own turning point, so $P_{2}=0$ but $P_{0}\neq 0$%
. We call it scenario TP2. It follows from (\ref{e1}) - (\ref{p2}) that

\begin{equation}
X_{2}=\frac{2\tilde{m}_{2}^{2}X_{0}}{\tilde{b}_{2}}\text{,}
\end{equation}%
\begin{equation}
X_{1}=\frac{X_{0}}{\tilde{b}_{2}}(\tilde{m}_{0}^{2}-\tilde{m}_{1}^{2}-\tilde{%
m}_{2}^{2})\text{,}
\end{equation}%
\begin{equation}
X_{0}=\frac{\tilde{b}_{2}N}{2\tilde{m}_{2}}\text{.}
\end{equation}

For ejection along the trajectory, the case under discussion is realized in
scenario II and we obtain from (\ref{P2f}), (\ref{P1II}) with $P_{2}=0$, $%
P_{0}\neq 0$ that%
\begin{equation}
X_{0}=\frac{b_{2}N}{2m_{2}}\text{.}  \label{xtn}
\end{equation}%
\begin{equation}
X_{2}=m_{2}N=\frac{2m_{2}^{2}}{b_{2}}X_{0}.  \label{x2tp2}
\end{equation}

Then, (\ref{x12}) gives us

\begin{equation}
X_{1}=\frac{X_{0}}{b_{2}}(m_{0}^{2}-m_{1}^{2}-m_{2}^{2})\text{.}
\end{equation}%
From (\ref{L2I}), (\ref{L1II}) we have%
\begin{equation}
L_{2}=0\text{, }L_{1}=L_{0}\text{.}
\end{equation}

One can check using (\ref{xtn}) and (\ref{x2tp2}) that we can also rewrite $%
X_{2}$ in the form 
\begin{equation}
X_{2}=\frac{b_{2}}{2m_{0}^{2}}X_{0}-\frac{\sqrt{d}}{2m_{0}^{2}}\sqrt{%
X_{0}^{2}-m_{0}^{2}N^{2}}\text{.}  \label{x2tp2d}
\end{equation}

\subsection{Case (i), turning point for particle 0 only (TP0)\label{tp0}}

In general scenario we take the limit $P_{0}\rightarrow 0$. This means that
decay occurs in the turning point of particle 0. For shortness, we call it
TP0. Particles 1 and 2 may have nonzero $P_{1}$ and $P_{2}$, so the point $%
r_{d}$ is not the turning point for them. It is clear from the conservation
law (\ref{P0}) that scenario TP0 can be realized in scenario I but not in
II. We obtain from (\ref{e1}) - (\ref{p2}) that%
\begin{equation}
P_{1}=P_{2}=\frac{N\sqrt{\tilde{d}}}{2\tilde{m}_{0}},
\end{equation}%
\begin{equation}
X_{1}=\frac{N}{2\tilde{m}_{0}}\tilde{b}_{1},
\end{equation}

\begin{equation}
X_{2}=\frac{N}{2\tilde{m}_{0}}\tilde{b}_{2}.
\end{equation}

Three quantities $E_{0}$, $L_{0}$, $N$ are related by one equation 
\begin{equation}
X_{0}=\tilde{m}_{0}N\text{.}  \label{t01}
\end{equation}

It follows from (\ref{t01}) immediately that 
\begin{equation}
L_{0}^{2}=r_{d}^{2}\frac{X_{0}^{2}-m_{0}^{2}N^{2}}{N^{2}}\text{.}  \label{t}
\end{equation}

In general, $L_{1}$ (or $L_{2}=L_{0}-L_{1})$ is a free parameter. It is only
restricted by the condition (\ref{Lminmsx})$.$

Let particles 1 and 2 be ejected in point $r_{d}$ along the trajectory of
particle 0. Now, it follows from (\ref{ej}) that either $L_{0}=0$ or $%
P_{1}=0=P_{2}$. By assumption, the second situation is now impossible and
will be considered in the next subsection. Now, we put $L_{0}=0$, whence $%
\tilde{m}_{0}=m_{0}$, so 
\begin{equation}
X_{0}=m_{0}N.\text{ }  \label{xon}
\end{equation}%
Then, it follows from (\ref{e2I}) - (\ref{L1I}) that 
\begin{equation}
X_{1}=X_{0}\frac{b_{1}}{2m_{0}^{2}}=\frac{b_{1}}{2m_{0}}N\text{,}
\end{equation}%
\begin{equation}
X_{2}=X_{0}\frac{b_{2}}{2m_{0}^{2}}=\frac{b_{2}}{2m_{0}}N\text{.}
\label{x20}
\end{equation}

\subsection{Case (ii), turning point for all three particles (TP3)}

Let us call this scenario TP3. Eq. It follows from (\ref{P}) that%
\begin{equation}
X_{0}=\tilde{m}_{0}N\text{,}  \label{x0n}
\end{equation}%
\begin{equation}
X_{1}=\tilde{m}_{1}N\text{, }
\end{equation}%
\begin{equation}
X_{2}=\tilde{m}_{2}N\text{.}
\end{equation}

With (\ref{x12}) taken into account, this leads to

\begin{equation}
\tilde{m}_{0}=\tilde{m}_{1}+\tilde{m}_{2}\text{.}  \label{m12}
\end{equation}

Then, after some algebraic manipulations, we can again obtain the same
expressions for $X_{1}$ and $X_{2}$ as in scenarios I and II, but with one
important difference. In scenario I, the expression for $X_{2}$ and,
correspondingly, energy $E_{2}$ for particle 2 that is enable to escape to
infinity, contained sign "minus" before the square root. This was due to the
necessity to have non-negative factor in (\ref{P2f}) inside parentheses.
Meanwhile, now this is irrelevant since $P_{2}=0$ due to the factor $P_{0}=0$%
. As a result, escaping particle 2 can have not only sign "minus" but also
sign "plus", so we are free to take%
\begin{equation}
E_{2}=q_{2}\varphi +\frac{b_{2}}{2m_{0}^{2}}X_{0}+\frac{\sqrt{d}}{2m_{0}^{2}}%
\sqrt{X_{0}^{2}-m_{0}^{2}N^{2}}\text{.}  \label{x12t}
\end{equation}%
If so, for particle 1 we have%
\begin{equation}
E_{1}=q_{1}\varphi +\frac{b_{1}}{2m_{0}^{2}}X_{0}-\frac{\sqrt{d}}{2m_{0}^{2}}%
\sqrt{X_{0}^{2}-m_{0}^{2}N^{2}}\text{.}  \label{1t}
\end{equation}

Equivalently,%
\begin{equation}
E_{2}=q_{2}\varphi +\frac{b_{2}}{2m_{0}^{2}}\tilde{m}_{0}N+\frac{\sqrt{d}%
\left\vert L_{0}\right\vert }{2m_{0}^{2}r_{d}}N\text{,}  \label{e2tp}
\end{equation}%
\begin{equation}
E_{1}=q_{1}\varphi +\frac{b_{1}}{2m_{0}^{2}}\tilde{m}_{0}N-\frac{\sqrt{d}%
\left\vert L_{0}\right\vert }{2m_{0}^{2}r_{d}}N\text{.}  \label{e1tp}
\end{equation}%
From (\ref{m12}) we have, taking the square and solving the quadratic
equation that

\begin{equation}
L_{2}=\frac{L_{0}b_{2}+signL_{0}\sqrt{d}\tilde{m}_{0}r_{d}}{2m_{0}^{2}}%
=L_{2}^{(+)},  \label{tp2}
\end{equation}

\begin{equation}
L_{1}=\frac{L_{0}b_{1}-signL_{0}\sqrt{d}\tilde{m}_{0}r_{d}}{2m_{0}^{2}}%
=L_{1}^{(-)}\text{,}  \label{tp1}
\end{equation}%
where $L^{(+)}$, $L^{(-)}$ are defined in (\ref{L+-}). Distinction between
particles 1 and 2 is almost conditional now but with the reservation that
the sign in the second term in (\ref{e2tp}) or in (\ref{e1tp}) correlates
with that in (\ref{tp2}), (\ref{tp1}).

In a sense, when from scenarios I and II we pass to scenario TP3, there is
an exchange of branches "plus" and "minus" between both particles in the
point where $P_{0}=0$ for non-radial motion. The aforementioned difference
disappears if $L_{0}=0$. Then, it follows from (\ref{x0n}) that $%
X_{0}=m_{0}N $. If also $d=0$, we return to the process described in \cite%
{conf}, \cite{myconf}. However, for nonzero $L_{0}$ the case under
discussion is more general.

When both $P_{0}\rightarrow 0$ and $\tilde{d}\rightarrow 0$, all scenarios
discussed in this section agree with each other.

\section{Threshold for the Penrose process\label{thres}}

Now, we are going to elucidate, when the Penrose process is possible. For
rotating black holes, this requires the existence of the ergoregion where,
by definition, $g_{00}>0$ \cite{pen}. In the RN case, $g_{00}$ does not
change the sign outside the horizon and there is no ergoregion in a usual
sense but there exists its analogue - so-called generalized ergosphere \cite%
{ruf}, \cite{den} where negative energy are allowed. It is sensitive to the
properties of particles. When $E_{1}<0\,$, the PP becomes possible and
particle 2 with an excess of energy goes to infinity.\ Meanwhile, in
scenario II both particles fall in a black hole. Therefore, we consider
scenario I$.$ The condition $E_{1}<0$ leads to 
\begin{equation}
X_{0}\frac{b_{1}}{2m_{0}^{2}}+q_{1}\varphi +\frac{\sqrt{d}\sqrt{%
X_{0}^{2}-m_{0}^{2}N^{2}}}{2m_{0}^{2}}\leq 0.
\end{equation}%
For $q_{1}\geq 0$ this is impossible. Let $q_{1}=-\left\vert
q_{1}\right\vert <0$. Then this condition takes the form%
\begin{equation}
2\left\vert q_{1}\right\vert m_{0}^{2}\varphi \geq X_{0}b_{1}+\sqrt{d}\sqrt{%
X_{0}^{2}-m_{0}^{2}N^{2}}.  \label{pp}
\end{equation}

Moreover, the SPP is now possible as well. Indeed, for a fixed $q_{0}$, the
energy $E_{2}=X_{2}+q_{2}\varphi $ formally grows unbounded when $%
q_{2}\rightarrow \infty $. The fact that the electromagnetic field can
significantly enhance the efficiency of the Penrose process was pointed in 
\cite{dad1}, \cite{dad2}. Meanwhile, for sufficiently high $q_{2}$ this is
quite generic feature that does not require the presence of a black hole.
This can happen even in the flat space-time \cite{flat}.

Let us consider an important particular case. If, in the framework of
scenario I, decay occurs in the turning point of particle 2, we can
substitute here (\ref{xtn}) in the point of decay and obtain%
\begin{equation}
E_{2}=q_{2}\varphi +Nm_{2}\text{,}  \label{e2n}
\end{equation}%
\begin{equation}
E_{1}=q_{1}\varphi +N\frac{(b_{1}b_{2}+d)}{4m_{0}^{2}m_{2}}=q_{1}\varphi +N%
\frac{m_{0}^{2}-m_{1}^{2}-m_{2}^{2}}{2m_{2}}.  \label{e1n}
\end{equation}%
It follows from (\ref{e1n}) and (\ref{b}), (\ref{d}) that in this case (\ref%
{pp}) can be rewritten in the form

\begin{equation}
\left\vert q_{1}\right\vert \varphi >N\frac{m_{0}^{2}-m_{1}^{2}-m_{2}^{2}}{%
2m_{2}}\text{.}  \label{ne}
\end{equation}

If all $q_{i}\rightarrow 0$, the results coincide with those obtained in the
static limit of a rotating black hole - see eq. 111 in \cite{ob}, \cite{wald}%
, eq. 3.30 in \cite{pani} and \cite{eq} and . Then, the Penrose process is
impossible as it should be. For $\left\vert q_{1}\right\vert \neq 0$, the PP
is possible in the scenario under discussion, if the point of decay,
according to (\ref{ne}), is located sufficiently close to the horizon.

In the particular case, when $d=0$, we have $m_{0}=m_{1}+m_{2}$ \cite{conf}
and the above condition turns into%
\begin{equation}
\left\vert q_{1}\right\vert \varphi >N_{d}m_{1}\text{,}  \label{th}
\end{equation}%
typical of the PP for pure radial motion, if decay occurs in the turning
point \cite{conf}.

Eq. (\ref{th}) can be interpreted as the statement that the electrostatic
energy of particle in an external fled should be bigger than the red-shifted
energy measured by a local observer. It is worth noting that in the case
under discussion $r_{d}$ is the turning point for particle 2, but not for
particles 0 and 1. In particular, eq. (\ref{m012}) is not satisfied in
general.

It is instructive to note, for comparison, that in the case of rotating
metrics extraction of energy in scenario I inside the ergoregion cannot be
realized at all (see \cite{eq}, Sec. VIA).

\section{Velocities and gamma factors for scenario I\label{vel}}

The above results are given in terms of particles' masses. Meanwhile, the
approach can be reformulated kinematically in terms of velocities. For
definiteness, let us consider scenario I, so $\sigma _{2}=+1$. Then, one can
check using (\ref{gai}) and (\ref{e2I}) the validity of the equation%
\begin{equation}
V_{2}=\frac{v_{2}-V_{0}}{1-v_{2}V_{0}}  \label{v2}
\end{equation}%
that is nothing but the Lorentz law of adding velocities. Now, the gamma
factor of relative motion for particles 0 and 2 $\gamma _{02}=-u_{0\mu
}u_{2}^{\mu }$, whence

\begin{equation}
m_{2}m_{0}\gamma _{02}=\frac{X_{0}X_{2}+P_{0}P_{2}}{N^{2}}-\frac{L_{0}L_{2}}{%
r^{2}}\text{.}  \label{ga}
\end{equation}%
In (\ref{ga}) we used eqs. of motion (\ref{td}) - (\ref{P}) and took into
account that both particles move in the opposite direction that gives us
sign "plus" in the numerator of eq. (\ref{ga}). Then, one obtains%
\begin{equation}
\gamma _{02}=\frac{b_{2}}{2m_{0}m_{2}}\text{, }v_{2}=\frac{\sqrt{d}}{b_{2}}%
\text{.}
\end{equation}%
Thus we can rewrite the formula for $X_{2}$ in the form%
\begin{equation}
X_{2}=\frac{m_{2}}{m_{0}}\gamma _{02}(X_{0}-v_{2}\sqrt{%
X_{0}^{2}-m_{0}^{2}N^{2}}).
\end{equation}%
In a similar manner, we find

\begin{equation}
m_{2}m_{0}\gamma _{01}=\frac{X_{0}X_{1}-P_{0}P_{1}}{N^{2}}-\frac{L_{0}L_{1}}{%
r^{2}},  \label{g01}
\end{equation}%
\begin{equation}
\gamma _{01}=\frac{b_{1}}{2m_{0}m_{1}}\text{, }v_{1}=\frac{\sqrt{d}}{b_{1}}%
\text{,}
\end{equation}%
\begin{equation}
V_{1}=\frac{v_{1}+V_{0}}{1+V_{1}V_{0}}\text{,}
\end{equation}%
\begin{equation}
X_{1}=\frac{m_{1}}{m_{0}}\gamma _{01}(X_{0}+v_{1}\sqrt{%
X_{0}^{2}-m_{0}^{2}N^{2}}).
\end{equation}

Here, $\gamma _{0i}$ ($i=1,2$) has the meaning of the standard Lorentz
factor of relative motion. For an individual particle we have%
\begin{equation}
X_{i}=m\gamma _{i}N\text{, }\gamma _{i}=\frac{1}{\sqrt{1-V_{i}^{2}}}\text{, }%
i=0,1,2\text{,}  \label{gai}
\end{equation}%
so%
\begin{equation}
V_{i}=\sqrt{1-\left( \frac{m_{i}N}{X_{i}}\right) ^{2}\text{,}}
\end{equation}%
$V_{i}$ is the velocity is measured in the stationary frame.

The expressions for the energy can be rewritten as 
\begin{equation}
E_{2}=q_{2}\varphi +\gamma _{02}\frac{m_{2}}{m_{0}}(X_{0}-v_{2}\sqrt{%
X_{0}^{2}-m_{0}^{2}N^{2}})\text{.}  \label{e2c}
\end{equation}%
\begin{equation}
E_{1}=q_{1}\varphi +\gamma _{01}\frac{m_{1}}{m_{0}}(X_{0}+v_{1}\sqrt{%
X_{0}^{2}-m_{0}^{2}N^{2}})\text{.}  \label{e1c}
\end{equation}

For the existence of the PP, the key restriction on the electric charge (\ref%
{pp}) is required. It can be written in terms of velocities%
\begin{equation}
\frac{m_{1}}{m_{0}}(X_{0}+v_{1}\sqrt{X_{0}^{2}-m_{0}^{2}N^{2}})<\left\vert
q_{1}\right\vert \varphi \sqrt{1-v_{1}^{2}}\text{,}  \label{pp1}
\end{equation}%
whence%
\begin{equation}
1+v_{1}V_{0}<\rho \sqrt{1-v_{1}^{2}}\text{, }\rho =\frac{m_{0}}{X_{0}m_{1}}%
\left\vert q_{1}\right\vert \varphi \text{. }  \label{vro}
\end{equation}%
Using (\ref{gai}), we can also write $\rho =\frac{\left\vert
q_{1}\right\vert \varphi }{m_{1}N\gamma _{0}}$. Taking the square of (\ref%
{vro}), we obtain%
\begin{equation}
v_{1}^{2}+\frac{2v_{1}V_{0}}{V_{0}^{2}+\rho ^{2}}+\frac{1-\rho ^{2}}{%
V_{0}^{2}+\rho ^{2}<0}=(v_{1}-v_{+})(v_{1}-v_{-})<0\text{,}
\end{equation}%
\begin{equation}
v_{+}=-\frac{V_{0}}{V_{0}^{2}+\rho ^{2}}+\frac{\rho }{V_{0}^{2}+\rho ^{2}}%
\sqrt{\rho ^{2}+V_{0}^{2}-1}\text{,}  \label{v+}
\end{equation}%
\begin{equation}
v_{-}=-\frac{V_{0}}{V_{0}^{2}+\rho ^{2}}-\frac{\rho }{V_{0}^{2}+\rho ^{2}}%
\sqrt{\rho ^{2}+V_{0}^{2}-1}<0,
\end{equation}%
whence%
\begin{equation}
v_{1}<v_{+}\text{.}  \label{vv+}
\end{equation}%
One can check easily that $v_{+}<1$.

The requirement $v_{+}\geq 0$ gives us%
\begin{equation}
\rho ^{2}\geq 1\text{.}  \label{ro}
\end{equation}%
In particular, for the Schwarzschild metric $\rho =0$, $v_{+}<0$ and the PP
is impossible as it should be.

Eq. (\ref{ro}) can be rewritten in the form%
\begin{equation}
\left\vert q_{1}\right\vert \varphi \geq \frac{X_{0}}{m_{0}}m_{1}\text{.}
\end{equation}

Thus in the present section we suggested description of decay and PP in
kinematic language by analogy with the rotating case \cite{wald}. However,
there is qualitative difference. For the existence of the PP, in the
rotating case the relative velocity between a new fragment and particle 0
should be quite high, the particle being ultrarelativistic \cite{wald}. In
our case, there is no restriction on velocity that can be arbitrarily low.
Moreover, instead of the lower bound typical of the rotating metric and
process with neutral particles, now there exists the upper bound (\ref{vv+}).

\section{Scenario B and the maximum of efficiency\label{inv}}

Both scenarios I and II imply that the parent particle 0 moves from infinity
towards a black hole. Meanwhile, of interest is also another situation when
particle 0 bounces back from its turning point and only afterwards decays to
particles 1 and 2. Particle 1 flies towards a back hole, particle 2 moves in
the outward direction. We call this scenario B (the first letter of the word
"bounce"). Then, the formulas for the energies read%
\begin{equation}
E_{2}=q_{2}\varphi +\gamma _{02}\frac{m_{2}}{m_{0}}(X_{0}+v_{2}\sqrt{%
X_{0}^{2}-m_{0}^{2}N^{2}})\text{,}  \label{2b}
\end{equation}%
\begin{equation}
E_{1}=q_{1}\varphi +\gamma _{01}\frac{m_{1}}{m_{0}}(X_{0}-v_{1}\sqrt{%
X_{0}^{2}-m_{0}^{2}N^{2}})\text{.}  \label{1b}
\end{equation}

If we assume that in the frame comoving with particle 0, it is particle 2
that moves outwardly but is drifted in the inward direction due to motion of
particle 0, then instead of (\ref{v2}) we have%
\begin{equation}
V_{2}=\frac{V_{0}+v_{2}}{1+v_{2}V_{0}}\text{.}  \label{V2-}
\end{equation}%
One can check using (\ref{2b}) and (\ref{gai}) that (\ref{V2-}) is indeed
satisfied.

For particle 1 now%
\begin{equation}
V_{1}=\frac{v_{1}-V_{0}}{1-V_{1}V_{0}}\text{,}
\end{equation}

Here, as before, particle 2 moves to infinity (now in the same direction as
particle 0) and particle 1 moves in the inward direction (opposite to
particle 0). However, now the signs before the radicals are opposite to
those in (\ref{e2c}), (\ref{e1c}). The condition for the PP gives us now 
\begin{equation}
\gamma _{01}\frac{m_{1}}{m_{0}}(X_{0}-v_{1}\sqrt{X_{0}^{2}-m_{0}^{2}N^{2}}%
)<\left\vert q_{1}\right\vert \varphi
\end{equation}%
instead of (\ref{pp1}). This entails%
\begin{equation}
v_{1}^{2}-\frac{2v_{1}V_{0}}{V_{0}^{2}+\rho ^{2}}+\frac{1-\rho ^{2}}{%
V_{0}^{2}+\rho ^{2}<0}=(v_{1}-v_{-})(v_{1}-v_{+})<0\text{,}
\end{equation}%
\begin{equation}
v_{\pm }=\frac{V_{0}}{V_{0}^{2}+\rho ^{2}}\pm \frac{\rho }{V_{0}^{2}+\rho
^{2}}\sqrt{\rho ^{2}+V_{0}^{2}-1}\text{.}
\end{equation}

It is easy to check that $v_{+}<1$. Here, there are two different cases.

If 
\begin{equation}
1-V_{0}^{2}\leq \rho ^{2}\leq 1,  \label{ro1}
\end{equation}%
then $v_{-}\geq 0$ and

\begin{equation}
v_{-}<v_{1}<v_{+}\text{.}
\end{equation}

If 
\begin{equation}
\rho ^{2}>1,  \label{ro+}
\end{equation}%
then $v_{-}<0$ and%
\begin{equation}
v_{1}<v_{+}\text{,}  \label{v1+}
\end{equation}%
so we have an upper bound on $v_{1}$.

If $\rho \rightarrow \infty $, $v_{+}\rightarrow 1$, so actually bound (\ref%
{v1+}) is satisfied automatically.

One can also choose TP3 as a scenario intermediate between I and B. Then,
again signs can be arranged according to (\ref{e2tp}), (\ref{e1tp}) with the
same conclusions about restrictions on the velocity of particle 1 that are
necessary for the PP to exist.

The significance of scenario B consists in that it enables us to attain the
maximum efficiency $\eta =\frac{E_{2}}{E_{0}}$ of the process. Indeed, in
this case the energy of particle 2 is given by eq. (\ref{2b}) with sign
"plus" before the radical.

It is seen from (\ref{x12t}) or (\ref{2b}) that $E_{2}$ is monotonically
decreasing function of $r$ since $\varphi =\frac{Qq_{2}}{r}$ with $q_{2}>0$
and $\frac{dN}{dr}>0$. Therefore, the most possible maximum of $E_{2}$ is
attained when decay happens near the horizon. One should bear in mind that
in this case $X_{0}$ should have the order $N_{d}$, so this is possible for
near-critical particle 0 only.

However, some important reservations about near-horizon decay are in order$.$
For a nonextremal black hole there is a potential barrier of a finite height
that prevents a near-critical particle with $X=(N_{d})$ in the near-horizon
region from approaching the horizon. Moreover, if a particle is exactly
critical, near the turning point $r_{t}$ from which a particle bounces back, 
$X^{2}=O(r_{t}-r_{+})^{2}$ near the horizon$.$ But for nonextremal black
holes $N^{2}\sim (r-r_{+})$ near the horizon, so $X^{2}\sim N^{4}$. As a
resut, it is seen from (\ref{P}) that the condition $P^{2}\geq 0$ cannot be
satisfied and a particle cannot penetrate into the near-horizon region at
all. If it is not exactly critical but near-critical, a particle is able to
move in the near-horizon region but such a particle can exist only between
the horizon and the turning point, so it cannot arrive from infinity anyway.
The situation is completely similar to that for rotating black holes \cite%
{gp}, \cite{gp42}, \cite{prd}, \cite{nearoz}. However, if a black hole is
extremal (or at least near-extremal) this becomes possible - see Sec. III of
Ref. \cite{rn} for more details about motion near turning points in the case
of the extremal RN black hole.

\subsection{Near-horizon limit for scenario B\label{nearb}}

It is worth noting what happens in the near-horizon limit $N\rightarrow 0$
within the scenario under discussion. On the first glance, it follows from (%
\ref{2b}) that $X_{2}\equiv E_{2}-q_{2}\varphi $ in this limit can be
arbitrary nonzero. However, this is not the case. The point is the
correlation between initial conditions and the type of a horizon. Let a
particle move away from the nonextremal horizon with finite nonzero $X_{2}$
(so-called usual particle). Then, if we continue its trajectory in the past,
it turns out that it appeared there some small proper time $\tau $ ago from
the region behind the horizon. But this would be a white, not a black hole
and is beyond the scope of our work. See also on details \cite{is} (there, $%
q=0$ but this does not matter in the context under discussion).

From the other hand, if the proper time required for crossing the horizon $%
\tau \rightarrow \infty $, such arguments do not work. This happens if the
horizon is extremal. Then, a particle can move in the outward direction from
the immediate vicinity of the horizon. But, in doing so, it must have $%
X=O(N) $ (see \cite{frontal} for details). Returning to our issue, we see
from (\ref{2b}) that both $X_{0}=O(N)$ and $X_{2}=O(N)$.

\section{Efficiency, proximity to horizon and type of scenario\label{prox}}

Usually, when considering decay, it is assumed that it happens in the
turning point of all particles. In doing so, it is often stated that the
efficiency reaches its maximum if decay occurs near the horizon (see, e.g.
the review \cite{fifty}). Meanwhile, these statements are not quite accurate
and require some essential reservations. Also, they apply to scenario TP3
but, in general, not to all other ones. As this concerns the important
aspects of process near a black hole, this needs more careful discussion
that is given below. Let us consider different types of scenarios in this
context case by case.

\subsection{Scenario I}

It is shown above that, according to (\ref{cr1}) this scenario requires $%
X_{0}\leq \frac{N_{d}b_{2}}{2m_{2}}$ in the point of decay $r_{d}.$ This
means that for a usual (not fine-tuned) particle that has $X_{0}\neq 0$ on
the horizon separated from zero, the horizon limit cannot be taken at all,
provided $m_{2}$ is a massive particle. If it has almost vanishing mass, $%
m_{2}=\mu _{2}N_{d}$ with $\mu _{2}\neq 0$, the situation changes since (\ref%
{cr1}) gives us%
\begin{equation}
X_{0}\leq \frac{b_{2}}{2\mu _{2}}.
\end{equation}

If this criterion is fulfilled, the horizon limit is indeed possible for a
usual particle. Otherwise, fine-tuning of particle 0 is required.

Let, for simplicity, particle 0 be neutral. Then, if%
\begin{equation}
q_{2}>q_{2}^{\ast }=\frac{\sqrt{M^{2}-Q^{2}}}{E_{0}Q}\sqrt{d}r_{+}\text{,}
\end{equation}%
the maximum is indeed achieved on the horizon (see \ref{app} for details).
At the same time, condition (\ref{pp}) should be fulfilled as well. On the
horizon it reduces to (\ref{thr}).

However, if $q_{2}<q_{2}^{\ast }$, the horizon corresponds not to the
maximum but to a local minimum of $E_{2}$.

If $m_{2}=0$ exactly, the horizon limit is also possible but according to (%
\ref{2m2}) this entails a quite strong condition $X_{2}=O(N_{d}^{2})$.

\subsection{Scenario II}

Here, there are no difficulties with decay near the horizon since inequality
(\ref{cr2}) can be satisfied easily. However, in this scenario both
particles fall in a black hole, so this option ceases to be \ "profitable".

\subsection{Scenario B}

Now, it follows from (\ref{2b}) that for any $q_{2}Q>0$, $\frac{\partial
E_{2}}{\partial r}<0$, so maximum is indeed achieved on the horizon.
However, another difficulty comes into play here. A usual particle cannot
start its motion near the horizon in the outward direction - see \ref{nearb}
above, Sec. IA in \cite{is} and \cite{frontal}. This is allowed for
fine-tuned or near-fine-tuned particles only, with $X_{2}=O(N_{d})$. The
same reasonings apply to scenario TP3.

\subsection{Scenario TP2}

According to (\ref{x2tp2}),

\begin{equation}
E_{2}=\frac{q_{2}Q}{r}+m_{2}N.
\end{equation}

Let us, for simplicity, consider the case of the extremal black hole, so $%
N=1-\frac{r_{+}}{r},Q=M=r_{+}$. Then,%
\begin{equation}
\frac{\partial E_{2}}{\partial r}=\frac{r_{+}}{r^{2}}(m_{2}-q_{2})\text{.}
\end{equation}

Thus the biggest value of $E_{2}$ is achieved on the horizon under the
condition $m_{2}<q_{2}$ only.

\subsection{Scenario TP0}

According to (\ref{x20}),%
\begin{equation}
E_{2}=\frac{q_{2}Q}{r}+\frac{b_{2}}{2m_{0}}N\text{.}
\end{equation}

Then, for the extremal black hole we have

\begin{equation}
\frac{\partial E_{2}}{\partial r}=\frac{r_{+}}{r^{2}}(\frac{b_{2}}{2m_{0}}%
-q_{2})\text{.}
\end{equation}

The horizon corresponds to the biggest value of $E_{2}$ for $q_{2}>\frac{%
b_{2}}{2m_{0}}$ only.

We see that one should be very careful making the statement about maximum of
efficiency. This necessarily includes indication of scenario and reservation
about relation between parameters.

\section{Different scenarios: comparison of properties\label{comp}}

Now, it is convenient to summarize the main features of all scenarios in
Table I.

\begin{tabular}{|l|l|l|l|l|l|}
\hline
Scenario & Particle 0 & Particle 2 & $\delta $ & $\left( v_{1}\right) _{\min
}$ mandatory & $X_{2}$ near horizon \\ \hline
I & in & out & $-1$ & no & $O(N)$ for $m_{2}\neq 0$, $O(N^{2})$ for $m_{2}=0$
\\ \hline
II & in & in & $+1$ & yes & $O(1)$ \\ \hline
TP2 & in & 0 & $-1$ & no & $O(N)$ \\ \hline
TP0 & 0 & out &  & no & $O(N)$ \\ \hline
TP3 & 0 & 0 & $\pm 1$ & no if $\delta =-1$, yes if $\delta =+1$ & $O(N)$ \\ 
\hline
B & out & out & $+1$ & no & $O(N)$ \\ \hline
\end{tabular}

\bigskip Table I. Classification and main features of scenarios.

We included in it scenarios in which particles 1 and 2 are ejected along the
trajectory of particle 0. However, this is done with one exception in a
degenerate case. In scenario TP0 not only $P_{0}=0$ but also $L_{0}=0$ - see
Sec. \ref{tp0}. Then, all the components of the three-momentum vanish, so
particle 0 is in rest in this point and there is no tangent vector to the
trajectory.

We can write a unifying formula for the energy of particle 2. If it moves in
the outward direction after decay of particle 0, it can, in principle,
escape to infinity and is potentially subject to the PP. Otherwise, it falls
in a black hole.%
\begin{equation}
E_{2}=q_{2}\varphi +X_{0}\frac{b_{2}}{2m_{0}^{2}}+\delta \frac{\sqrt{d}\sqrt{%
X_{0}^{2}-m_{0}^{2}N^{2}}}{2m_{0}^{2}}.  \label{e2d}
\end{equation}
If immediately after decay particle 2 moves in the outward/inward direction,
we use shortening "out"/"in". If it is in the radial turning point, we write
"0". In scenario TP0 the factor $\sqrt{X_{0}^{2}-m_{0}^{2}N^{2}}=0$, so $%
\delta $ is irrelevant.

The full trajectory of particle 2 is model-dependent and cannot be found
without specifying the metric. The efficiency $\eta =\frac{E_{2}}{E_{0}}$.
In the 4th column we indicate whether or not some low bound $v_{1}\geq
\left( v_{1}\right) _{\min }>0$ is required for the PP to occur. In scenario
II particle 2 falls in a black hole, so the low bound on $v_{1}$ is pure
formal since it has nothing to do with the PP. The general feature consists
in that there is such a bound for $\delta =+1$ and it is absent if $\delta
=-1$.

In principle, a combined scenario is also possible. If in a point of decay
both particles moves in the inward direction (scenario II) but after
bouncing back in the turning point particle 2 changes direction and moves
outwardly. In particular, this can happen in the near-horizon region of the
extremal black hole with $X_{2}=O(N)$ \cite{rn}, \cite{fh}.

In all cases, formally $E_{2}\rightarrow \infty $ when $q_{2}\rightarrow
\infty $. However, in realistic situations $q_{2}$ is bounded \cite{rn}, 
\cite{fh}.

\section{Discrepancy with two previous works on the subject\label{discr}}

The Penrose process with charged particles was also considered in \cite{ju}
where an additional assumption that the black hole charge $Q$ is small was
made. Our results do not agree. The authors of \cite{ju}, according to their
eqs. (34), (35), obtained that the outgoing particle after decay near the
horizon has a finite nonzero energy. Meanwhile, it follows from our formulas
that in this case the energy of an escaping particle (if we neglect $Q$)
tends to zero. If we take the charge $Q$ into account, in a similar way the
difference $X_{2}=E_{2}-q_{2}\frac{Q}{r_{+}}$ tends to zero. Actually, the
subtlety in the issue under discussion consists in the necessity to take
into account correlation between dynamics and kinematics. This means that in
scenario I (where particle 2 escapes) the sign before the radical in (\ref%
{e2I}) comes with minus, not plus. Instead, one can consider scenario II
with finite $X_{1}$ and $X_{2}$ where one of particles has the sign plus but
this particle falls in a black hole and does not escape. These features can
be seen in Table I.

One more attempt of considering the PP for charged particles was made in 
\cite{sh} with a magnetic field $B$ taken into account. In our view, the
results for efficiency $\eta =\frac{E_{2}}{E_{0}}$ (in our notations)
described by Eqs. (31) and (32) of Ref. \cite{sh} are incorrect. In the
neutral case $q=0$, $Q=0$ and $B=0$ they give $\eta =0$ instead of the
Schwarzschild value $\eta _{Sch}$. Therefore, for the RN metric, it also
does not reproduce the formulas like (\ref{x12t}) of the present paper.

The reason of discrepancy can be explained as follows. The results (31),
(32) of \cite{sh} are based on their eq. (30) that contains the angular
velocity of a particle. Therefore, account for angular momentum is
necessary.\ For a decay, say, in the turning point of radial motion for all
three particles, the correct values are given by our eqs. (\ref{tp2}), (\ref%
{tp1}). Meanwhile, in \cite{sh} all angular momenta are put to be zero. This
leads to contradiction.

Alternatively, one may consider a scenario in which all particles move
radially with $L_{0}=L_{1}=L_{2}=0$. However, in this case eqs. (19), (20)
of \cite{sh} from which (31) and (32) were derived, loose their sense since
(19) is obtained for motion along a circle. Again, we obtain contradiction.

\section{Relation to the BSW effect\label{bsw}}

Up to now, we discussed the standard Penrose process based on particle
decay. Meanwhile, the aforementioned properties are applicable to the
collisional version of the PP as well, when particles 1 and 2 collide to
produce particles 3 and 4. This is due to the fact that particles 1 and can
be considered as a combined one with characteristics obeying (\ref{e}) - (%
\ref{x12}) and $m_{0}$ equal to the energy $E_{c.m.}$ in the center of mass
frame \cite{tz}. The BSW process and properties of debris were considered
before in somewhat different approach in which consideration was restricted
from the very beginning to the immediate vicinity of the horizon and
approximate formulas were used \cite{upper}. Now, it is instructive to
compare it with the present approach when one starts from exact formulas
from the very beginning.

We are mainly interested in the situation when collisions lead to the BSW
effect. To make comparison possible, we assume that the RN black hole is
extremal and consider pure radial motion like in \cite{rn}. For the extremal
RN black hole, the Coulomb potential $\varphi =1-N$. Let particle 1 be
critical. By definition, this means that on the horizon $X=0$. Then, in this
setting, $X_{1}=E_{1}N$. We want to show that our exact formulas in the
limit when $N_{c}\rightarrow 0$ turn into the results of \cite{rn}. \ (Here
subscript "c" denotes the point of collision.) It is sufficient to trace
correspondence with eq. (26) of \cite{rn} which is the key point of the
analysis there.

According to the BSW effect \cite{ban} and its electric counterpart \cite{jl}%
, the energy $E_{c.m.}$ in the center of mass of two colliding particles
grows unbounded if (i) one of particles (say, 1) is critical, (ii) collision
happens near the horizon, when $N\rightarrow 0$. In our context,%
\begin{equation}
m_{0}^{2}\approx \frac{\beta }{N}\text{,}
\end{equation}%
where $\beta $ is a constant \cite{centr} equal to%
\begin{equation}
\beta =2\left( X_{2}\right) _{c}A\text{, }A=E_{1}-\sqrt{E_{1}^{2}-m_{1}^{2}}%
\text{.}
\end{equation}%
This formula can be obtained in the near-horizon limit directly from (\ref%
{ga}), if it is applied to particles 1 and 2. As a result,%
\begin{equation}
\sqrt{X_{0}^{2}-m_{0}^{2}N^{2}}\approx X_{0}-\frac{\beta N}{2X_{0}}\text{,}
\end{equation}%
where $X_{0}=X_{2}+E_{1}N\approx X_{2}$ near the horizon.

Particle 4 falls in a black hole, particle 3 is near-critical particle.
Similarly to \cite{rn}, we define 
\begin{equation}
q_{3}=E_{3}(1+\delta )\text{,}
\end{equation}%
where%
\begin{equation}
\delta =CN_{c}\ll 1\text{,}
\end{equation}%
subscript "c" means the point of collision, $C$ is a constant.

Then, $X_{3}$ and $X_{4}$ are given by our formulas in which 1 should be
replaced by 4 and 2 should be replaced by 3. \ We have%
\begin{equation}
X_{3}=E_{3}-q_{3}+q_{3}N=E_{3}(N-CN_{c})\text{,}
\end{equation}%
\begin{equation}
X_{3}(N_{c})=E_{3}N_{c}(1-C)\text{,}
\end{equation}%
\begin{equation}
P_{3}(N_{c})=N_{c}\sqrt{E_{3}^{2}(1-C)^{2}-\beta }\text{.}
\end{equation}%
From the other hand, it follows from (\ref{g01}) with $\gamma _{01}$
replaced by $\gamma \equiv \gamma _{12}$ that%
\begin{equation}
\gamma m_{1}m_{2}\approx \frac{\left( X_{2}\right) _{c}(E_{1}-\sqrt{%
E_{1}^{2}-m_{1}^{2}})}{N_{c}}\text{,}
\end{equation}%
\begin{equation}
b_{3}=m_{0}^{2}+m_{3}^{2}-m_{4}^{2}\text{,}
\end{equation}%
\begin{equation}
d=b_{3}^{2}-4m_{0}^{2}m_{3}^{2}.
\end{equation}%
Taking into account (\ref{e2I}) for $N_{c}\ll 1$, we obtain%
\begin{equation}
E_{3}(1-C)=\frac{m_{3}^{2}}{2A}+\frac{1}{2}A\text{.}
\end{equation}

This agrees completely with eq. (26) \cite{rn}, so further analysis from the
aforementioned paper applies. See also \cite{upper}. If particle 1 moves
from the vicinity of the horizon in the outward direction (so-called
Schnittman process \cite{revised}), consideration runs along the same lines.

\section{ Summary and Conclusions\label{concl}}

Thus we considered the Penrose process for motion of all three particles
within the same plane. We relied on exact formulas for characteristics of
daughter particles in terms of a parent one. In doing so, we gave full
classification of possible scenarios and derive the bounds on possible
values of angular momenta of daughter particles. We selected those scenarios
for which ejection occurs along the trajectory of a parent particle.
Significance of scenarios of such a type consists in that we obtain maximum
(minimum) value of energy. This is confirmed by direct computation of
extrema of the energy of corresponding particles with respect to the angular
momentum of one of them.

We formulated the results in two forms. The first one includes the
dependence of the outcome on masses of particles. The second form expresses
them in terms of velocities and gamma factors of relative motion between a
parent particle and daughter ones. The corresponding formulas obtained for a
static charged metric are similar to whose obtained in \cite{wald} for
rotating metrics and neutral particles. Meanwhile, there are some essential
differences. For the Penrose process to occur in the Kerr or other rotating
metric, there is a lower bound on velocity of a daughter particle in the
center of mass frame, this bound being rather high \cite{wald}. This creates
obstacles for using this process. Meanwhile, in our case, the situation is
more diverse and there are scenarios in which the velocity obeys the upper
(not lower)\ bound, so the Penrose process can be realized more easily.

We discussed in detail an important question how the efficiency of the
Penrose process depends on a point in which decay occurred. The results
depends in general on a scenario and in this sense classification of
scenarios developed in our work enables to elucidate this issue in general
setting. One may think that this will be useful for analysis of diverse
process near astrophysical black holes since we did not restrict ourselves
by a separate scenario or special set of date.

Although we discussed decay, the results are applicable to other reactions
between particles including the collisional Penrose process in which
particles 1 and 2 collide to create new particles 3 and 4. In particular,
this can lead to the BSW effect. We traced how the current approach based on
exact formulas agrees with the previous one used before for description of
the aforementioned effect.

In our consideration, we implied that the metric is the Reissner-Nordstr\"{o}%
m one. However, the approach can be applied to a more general metrics of
type (\ref{met}).

Our goal consisted in the present article not in considering some concrete
astrophysical problem but in development of general formalism. We carried
out the analysis of possible types of scenarios in the simplest case of
processes with charged particles thus developing general approach. This
implied consideration of a static metric and electric field. The next step
is supposed to be inclusion of rotation, electric charge and a magnetic
field altogether into consideration.

We hope that the corresponding formalism will be useful for solving
realistic problems in astrophysics. This is supposed to include processes in
the accretion discs, extraction of energy from magnetized black holes,
properties of ionization of neutral particle falling into a black hole, etc.
This needs separate treatment. We hope that model-independent approach
developed in our previous work \cite{eq} and the present one, will be useful
for a diverse set of such problems.

\section{Appendix: dependence of efficiency on a point\label{app}}

For simplicity, we consider here a neutral particle 0. Then, $q_{0}=0$, $%
q_{1}=-q_{2}$ and $X_{0}=E_{0}$. Let after decay the energy $E_{2}$ (and
thus efficiency $\eta =E_{2}/E_{0})$ is given by eq. of the type 
\begin{equation}
E_{2}=q_{2}\frac{Q}{r}+\frac{b_{2}}{2m_{0}^{2}}E_{0}+\frac{\sqrt{d}}{%
2m_{0}^{2}}\sqrt{E_{0}^{2}-m_{0}^{2}N^{2}}\text{.}
\end{equation}

Here, $q_{2}>0$. As $\frac{dN}{dr}>0$, it follows that $\frac{dE_{2}}{dr}>0$
and the maximum is formally achieved for decay on the horizon but with all
necessary reservations made in Sec. \ref{prox} above. More involved case
arises if%
\begin{equation}
E_{2}=q_{2}\frac{Q}{r}+\frac{b_{2}}{2m_{0}^{2}}E_{0}-\frac{\sqrt{d}}{%
2m_{0}^{2}}\sqrt{E_{0}^{2}-m_{0}^{2}N^{2}}\text{.}
\end{equation}

Then, direct calculation shows that on the horizon%
\begin{equation}
\left( \frac{dE_{2}}{dr}\right) _{+}=\frac{1}{r_{+}^{2}}(\frac{\sqrt{%
M^{2}-Q^{2}}}{E_{0}}\sqrt{d}-q_{2}Q)\text{.}
\end{equation}

Thus if the biggest value of $E_{2}$ is required to occur on the horizon, it
is necessary that $\left( \frac{dE_{2}}{dr}\right) _{+}<0$, whence%
\begin{equation}
q_{2}>q_{2}^{\ast }=\frac{\sqrt{M^{2}-Q^{2}}}{E_{0}Q}\sqrt{d}.
\end{equation}

However, if 
\begin{equation}
q_{2}<q_{2}^{\ast }\text{,}  \label{less}
\end{equation}%
the horizn crresponds to the smallest value.

Simultaneously, the threshold for the PP (\ref{pp}) gives us on the horizon
for $\left\vert q_{1}\right\vert =q_{2}$%
\begin{equation}
q_{2}\geq \frac{E_{0}r_{+}}{2m_{0}^{2}Q}(b_{1}+\sqrt{d}),  \label{thr}
\end{equation}%
where we have taken into account that now $\left\vert q_{1}\right\vert
=q_{2} $. Both inequalities (\ref{less}) and (\ref{thr}) are compatible with
each other, provided%
\begin{equation}
\frac{E_{0}^{2}}{m_{0}^{2}}\leq 2\frac{\sqrt{d}\sqrt{M^{2}-Q^{2}}}{(b_{1}+%
\sqrt{d})r_{+}}.
\end{equation}


\begin{thebibliography}{99}
\bibitem{pen} R. Penrose, Gravitational collapse: The role of general
relativity. Riv. Nuovo Cimen. \textbf{1,} 252 (1969).

\bibitem{pen2} R. Penrose, R. M. Floyd, Extraction of Rotational energy from
a Black Hole, Nature \textbf{229}, 177 (1971).

\bibitem{ruf} G. Denardo and R. Ruffini, On the energetics of Reissner
Nordstr\"{o}m geometries, Phys. Lett. \textbf{B} 45, 259 (1973).

\bibitem{den} G. Denardo, L. Hively, and R. Ruffini, On the generalized
ergosphere of the Kerr-Newman geometry, Phys. Lett. B \textbf{50, }270
(1974).

\bibitem{schrev} J. D. Schnittman, The collisional Penrose process, Gen
Relativ Gravit \textbf{50}, 77 (2018), [arXiv:1910.02800].

\bibitem{conf} T. Kokubu , S.-L. Li, P. Wu, and Hongwei Yu. Confined Penrose
process with charged particles, Phys. Rev. D. \textbf{104}, 104047 (2021),
[arXiv:2108.13768].

\bibitem{myconf} O. B. Zaslavskii, Confined Penrose process and black-hole
bomb Phys. Rev. D \textbf{106}, 024037 (2022), [arXiv:2204.12405].

\bibitem{ban} M. Ba\~{n}ados, J. Silk and S.M. West, Kerr black holes as
particle accelerators to arbitrarily high energy, Phys. Rev. Lett. \textbf{%
103}, 111102 (2009) [arXiv:0909.0169].

\bibitem{ext} M. Zhang, J. Jiang, Y. Liu, and W-B. Liu, Collisional Penrose
process of charged spinning particles, Physical Review D \textbf{98}, 044006
(2018).

\bibitem{bin} L. T. Sanches, and M. Richartz, Energy extraction from
non-coalescing black hole binaries, Physical Review D \textbf{104}, 124025
(2021), [arXiv:2108.12408].

\bibitem{pp2bh} A. Baez, N. Breton , and I. Cabrera-Munguia, Energy
extraction in electrostatic extreme binary black holes, Physical Review D 
\textbf{106}, [arXiv:2210.08193].

\bibitem{ppv} V. Vertogradov, Extraction energy from charged Vaidya black
hole via the Penrose process, Commun. Theor. Phys. \textbf{75}, 045404,
[arXiv:2210.04784].

\bibitem{is} O. B. Zaslavskii, Is the super-Penrose process possible near
black holes? Phys. Rev. D 93 (2016), 024056 [arXiv:1511.07501].

\bibitem{trade} N. Tsukamoto, Is there a trade-off relation between
efficiency and power in a collisional Penrose process in an extreme
Reissner-Nordstr\"{o}m spacetime? Physical Review D \textbf{105, }104065
(2022), [arXiv:2205.02628].

\bibitem{centr} O. B. Zaslavskii, Center of mass energy of colliding
electrically neutral particles and super-Penrose process, Phys. Rev. D 
\textbf{100}, 024050 (2019) [arXiv:1904.04874].

\bibitem{rn} O. B. Zaslavskii, Energy extraction from extremal charged black
holes due to the BSW effect. Phys. Rev. D\textbf{\ 86}, 124039 (2012)
[arXiv:1207.5209].

\bibitem{fh} \qquad F. Hejda, J. P. S. Lemos, O. B. Zaslavskii, Extraction
of energy from an extremal rotating electrovacuum black hole: Particle
collisions in the equatorial plane, Phys. Rev. D \textbf{105}, 024014
(2022), [arXiv:2109.04477].

\bibitem{dad3} S. M. Wagh and N. Dadhich, The energetics of black holes in
electromagnetic fields by the Penrose process, Phys. Repts. \textbf{183,}
137 (1989).

\bibitem{rufkerr} J. A. Rueda, R. Ruffini, Extracting the energy and angular
momentum of a Kerr black hole, ur. Phys. J. C 83, 960 (2023)
[arXiv:2303.07760].

\bibitem{dad1} M. Bhat, S. Dhurandhar and N. Dadhich, Energetics of the
Kerr-Newman black hole by the Penrose process, J. Astrophys. Astron. \textbf{%
6,} 85 (1985).

\bibitem{dad2} S. Parthasarathy, S. M. Wagh, S. V. Dhurandhar and N.
Dadhich, High efficiency of the Penrose process of energy extraction from
rotating black holes immersed in electromagnetic fields, Astr. J., Part 1 
\textbf{307,} 38 (1986).

\bibitem{luca} L. Comisso, and F. A. Asenjo, Magnetic reconnection as a
mechanism for energy extraction from rotating black holes, Phys. Rev. D 
\textbf{103}, 023014 (2021), [arXiv:2012.00879].

\bibitem{fifty} A. Tursunov and N. Dadhich, Fifty years of energy extraction
from rotating black hole: revisiting magnetic Penrose process, Universe 
\textbf{5}, 125, (2019), [arXiv:1905.05321].

\bibitem{uniform} K. Gupta, Y. T. A. Law, and J. Levin, Penrose process for
a charged black hole in a uniform magnetic field, Phys. Rev. \textbf{D} 104,
084059 (2021), [arXiv:2106.15010].

\bibitem{win} V. Patel, K. Acharya, P. Bambhaniya, and P. S. Joshi, Energy
extraction from Janis-Newman-Winicour naked singularity, Phys. Rev. D 
\textbf{107}, 064036 (2023), [arXiv:2301.11052].

\bibitem{eq} O.B. Zaslavskii, On general properties of the Penrose process
with neutral particles in the equatorial plane, Phys. Rev. D 108, 084022
(2023), [arXiv:2307.06469].

\bibitem{ju} A. Tursunov, B. Juraev, Z. Stuchl\'{\i}k, M. Kolo\v{s},
Electric Penrose process: high-energy acceleration of ionized particles by
non-rotating weakly charged black hole, Phys. Rev. D \textbf{104}, 084099
(2021), [arXiv:2109.10288].

\bibitem{dad0} S. M. Wagh, S. V. Dhurandhar, and N. Dadhich, Revival of the
Penrose process for astrophysical applications, The Astrophysical Journal, 
\textbf{290}, 12 (1985).

\bibitem{wald} R. M. Wald, Energy limits on the Penrose process, Astrophys.
J. \textbf{191}, 231 (1974).

\bibitem{ob} Yu. V. Pavlov, O. B. Zaslavskii, Int. J. Mod. Phys. D \textbf{32%
}, 2250143 (2023), Particle decay, Oberth effect and a relativistic rocket
in the Schwarzschild background, [arXiv:2111.09240].

\bibitem{sh} Shaymatov S., Sheoran, P., Becerril R., Nucamendi U., Ahmedov
B., Efficiency of Penrose process in spacetime of axially symmetric
magnetized Reissner-Nordstr\"{o}m black hole, Phys. Rev. D. \textbf{106},
024039 (2022) [arXiv:2204.02026].

\bibitem{gp} A.A. Grib and Yu.V. Pavlov, On particles collisions in the
vicinity of rotating black holes, JETP\ Letters \textbf{92}, 125 (2010).

\bibitem{gp42} A. A. Grib and Yu.V. Pavlov, On Particle Collisions near
rotating black holes, Gravitation and Cosmology \textbf{17}, 42 (2011)
[arXiv:1010.2052].

\bibitem{prd} O.B. Zaslavskii, Acceleration of particles as universal
property of rotating black holes, Phys. Rev.\textit{\ }D\textbf{\ 82} (2010)
083004 [arXiv:1007.3678].

\bibitem{nearoz} H.\thinspace V. Ovcharenko and O.\thinspace B. Zaslavskii,
BSW phenomenon for near-fine-tuned particles with external force: general
classification of scenarios, [arXiv:2402.17383].

\bibitem{frontal} H.\thinspace V. Ovcharenko and O.\thinspace B. Zaslavskii,
High energy head-on particle collisions near event horizons: classifcation
of scenarios [arXiv:2404.03364].

\bibitem{tz} I. V. Tanatarov, O. B. Zaslavskii, Collisional super-Penrose
process and Wald inequalities, Gen Relativ Gravit \textbf{49} 119 (2017),
[arXiv:1611.05912].

\bibitem{flat} O. B. Zaslavskii, Pure electric Penrose and super-Penrose
processes in the flat space-time, Int. J. Mod. Phys. D. \textbf{28,} 1950062
(2019) [arXiv:1807.05763].

\bibitem{pani} R. Brito, P. Pani, V. Cardoso, Superradiance. New Frontiers
in Black Hole Physics. 2nd edition. Lecture Notes in Physics. Springer. 2015.

\bibitem{upper} T. Harada, H. Nemoto, and U. Miyamoto, Upper limits of
particle emission from high-energy collision and reaction near a maximally
rotating Kerr black hole, Phys. Rev. D 86, 024027 (2012); 86, 069902(E)
(2012).

\bibitem{jl} O. Zaslavskii, Acceleration of particles by nonrotating charged
black holes. JETP Letters \textbf{92}, 571 (2010), [arXiv:1007.4598].

\bibitem{revised} J. D. Schnittman, Revised upper limit to energy extraction
from a Kerr black hole, Phys. Rev. Lett. \textbf{113}, 261102 (2014),
[arXiv:1410.6446].
\end{thebibliography}
\end{document}